\documentclass[]{aastex63}
\received{}
\revised{}
\accepted{May 6, 2021}

\submitjournal{ApJL}

\shorttitle{The rotation curve of Elias 2-27}
\shortauthors{Veronesi al.}

\begin{document}

\title{A dynamical measurement of the disk mass in Elias 2-27}
%Deviation from Keplerian rotation in Elias 2-27: evidence of a massive disk?}

\correspondingauthor{Benedetta Veronesi}
\email{benedetta.veronesi@unimi.it}

\author[0000-0002-0786-7307]{Benedetta Veronesi}
\affiliation{Dipartimento di Fisica, Università degli Studi di Milano, Via Celoria 16, Milano, 20133, Italy}
\author[0000-0002-4044-8016]{Teresa Paneque-Carre\~no}
\affiliation{European Southern Observatory, Karl-Shwarzschild-Strasse 2, D-85748 Garching bei Munchen, Germany}
\author[0000-0002-2357-7692]{Giuseppe Lodato}
\affiliation{Dipartimento di Fisica, Università degli Studi di Milano, Via Celoria 16, Milano, 20133, Italy}
\author[0000-0003-1859-3070]{Leonardo Testi}
\affiliation{European Southern Observatory, Karl-Shwarzschild-Strasse 2, D-85748 Garching bei Munchen, Germany}
\author[0000-0002-1199-9564]{Laura M. P\'erez} \affiliation{Departamento de Astronom\'ia, Universidad de Chile, Camino El Observatorio 1515, Las Condes, Santiago, Chile}
\author{Giuseppe Bertin}
\affiliation{Dipartimento di Fisica, Università degli Studi di Milano, Via Celoria 16, Milano, 20133, Italy}
\author[0000-0002-8138-0425]{Cassandra Hall}
\affiliation{Department of Physics and Astronomy, The University of Georgia, Athens, GA 30602, USA.}
\affiliation{Center for Simulational Physics, The University of Georgia, Athens, GA 30602, USA.}

\begin{abstract}
Recent multi-wavelength ALMA observations of the protoplanetary disk orbiting around Elias 2-27 revealed a two armed spiral structure. The observed morphology together with the young age of the star and the disk-to-star mass ratio estimated from dust continuum emission make this system a perfect laboratory to investigate the role of self-gravity in the early phases of star formation. This is particularly interesting if we consider that gravitational instabilities could be a fundamental first step for the formation of planetesimals and planets. In this Letter, we model the rotation curve obtained by CO data of Elias 2-27 with a theoretical rotation curve including both the disk self-gravity and the star contribution to the gravitational potential. We compare this model with a purely Keplerian one and with a simple power-law function. We find that (especially for the $^{13}$CO isotopologue) the rotation curve is better described by considering not only the star, but also the disk self-gravity. We are thus able to obtain for the first time a dynamical estimate of the disk mass of $0.08\pm0.04\,M_{\odot}$ and the star mass of $0.46\pm0.03\,M_{\odot}$ (in the more general case), the latter being comparable with previous estimates. From these values, we derive that the disk is 17$\%$ of the star mass, meaning that it could be prone to gravitational instabilities. This result would strongly support the hypothesis that the two spiral arms are generated by gravitational instabilities.
\end{abstract}


\section{Introduction} \label{sec:intro}

Protoplanetary disks form in the chaotic environment of molecular cloud cores and in their early stages they are massive enough to have a non negligible effect on the evolution of the overall system. The disk self-gravity may influence the disk dynamics through the propagation of density waves that lead to the formation of prominent structures in the form of one or more spiral arms. These morphologies have been detected by ALMA and VLT-SPHERE in both Class 0/I and Class II systems, and they are usually assumed to be originated by embedded companions (e.g. HD135344B, \citealt{veronesi19}, MWC 758, \citealt{calcino20}) or by self-gravity  (e.g. Elias 2-27, \citealt{perezelias227,huang18c}). In the second case, density waves are thought to provide a non negligible contribution to the angular momentum transport and may have a crucial role in the formation of planetesimals through dust trapping at the location of the spirals and the following direct fragmentation of spiral overdensities into bound objects \citep{rice04,rice06,kratter16}. Being able to give an estimate of the disk mass is the first step in order to put other pieces in the puzzle of planet formation and to understand the origin of the observed spirals \citep{marel20,veronesi19,bergin18a}. But how can we determine the mass of these systems?

First, dust masses are typically inferred using the optically thin approximation at millimeter wavelengths. It is worth noting that, although it may be trivial, this estimate still carries a high level of uncertainty, due to the assumed optical depth of the dust at (sub-)mm wavelengths (e.g. the dust opacity and the level of dust growth, \citealt{bergin18a}). Once the dust mass is known, one needs to convert this into a total disk mass by assuming some gas/dust ratio, which is generally assumed to be equal to 100 \citep{draine03}, although this number is highly uncertain (see e.g. \citealt{macias21}). On the other hand, it is more difficult to quantify the disk mass from direct gas tracers. A common procedure is to use CO observations in its various isotopologues (such as $^{13}$CO and C$^{18}$O) as a proxy for the gas mass. But since the conversion of the observed CO mass into total gas mass is not well understood \citep{williams14a,bergin18a}, this is not straightforward. Another issue adding complexity to the problem is that the properties of different molecules also vary spatially and temporally, depending on the models \citep{ilee17,quenard18}. Indeed, estimates derived from CO observations result in very low disk masses compared to dust estimates \citep{pascucci16a,ansdell16a,miotello17,long17a}. \cite{miotello16a} associated this trend with multiple possible processes: carbon depletion in the disk  \citep{favre13,bosman18,cleeves18}, photodissociation in the upper layers, freeze-out at the disk midplane or in general other isotope-selective processes. Another aspect that should be considered comes from far-IR HD lines \citep{bergin13,trapman17} observations, suggesting that the gas-to-dust ratio measurement is affected by the fact that the emitting regions of various gas tracers differ from each other and in turn differ from the regions where the dust is observed.

We can also estimate the total disk mass in a dynamical way, by using the disk rotation curve, and detecting deviations from the expected Keplerian curve. This method has been widely used with galaxies \citep{barbieri05} and sometimes it has been used to estimate also the mass of AGN disks \citep{lodato03}.  
Usually protoplanetary disks are assumed to be Keplerian, since typically the stellar mass dominates over the disk mass, and their rotation curve can be sufficiently well described by the stellar contribution alone.
%Usually protoplanetary disks are assumed to be Keplerian, meaning that their rotation curve is determined by the star alone. 
Instead, when the disk contribution is significant, we could be able to fit the observed rotation curve and to give an independent disk mass estimate \citep{bertin99}. 
For relatively massive disks, this dynamical estimate is now possible since we have access to a large amount of gas kinematic data with high (angular and velocity) resolution and high sensitivity. From these data, we can infer the geometry of the disk and recover the height and the velocity of the emitting gas layer \citep{pinte18}.

One of the most interesting observed spiral structures is the one hosted by the protoplanetary disk orbiting around Elias 2-27. Elias 2-27 is a young 0.8 Myr M0 star \citep{andrews09} located at a distance of $\sim$115 pc \citep{gaia18} in the Ophiucus star-forming region \citep{luhman99}. The surrounding disk is unusually large and massive, with a disk-to-star mass ratio of $\sim$ 0.3 \citep{andrews09,perezelias227}, as estimated by converting dust mass into total disk mass with the usual gas/dust ratio of 100. ALMA observations of this system detected two large-scale spiral arms \citep{perezelias227}, which have been confirmed in the DSHARP survey at higher resolution \citep{andrews18a}. Together with these spiral arms, a 14 au wide, inner gap, located at 69 au from the star \citep{huang18b,huang18c} has been observed. Recent studies have confirmed that a possible origin for the spiral arms is the development of gravitational instabilities \citep{paneque20,hall20,halletal2018,forgan18a,meruetal2017,bae18b}. However, this physical mechanism does not explain the origin of the dust gap, which could have been carved by a companion of $\sim$0.1 $M_{\rm Jup}$ as constrained from hydrodynamical simulations by \cite{zhang18}. Moreover, localized deviations from Keplerian motions at the location of this dust gap have been found recently, reinforcing the hypothesis of a planetary-mass companion \citep{pinte20}. Yet, it has been shown that a low-mass inner companion would be able to explain the gap but not the origin of the observed spiral arms \citep{meruetal2017}.
With this background in mind, we decided to take a closer look at the rotation curve of this system in order to provide a dynamical mass estimate of the disk independent of dust-CO measurements and to test the viability of gravitational instabilities as the origin of the observed grand-design spiral structure. 

In this paper we study the rotation curve of the protoplanetary disk orbiting around Elias 2-27 by comparing two competing models: a Keplerian disk model and a self-gravitating disk model \citep{bertin99,lodato03}. The gravitational field has been computed by solving the Poisson equation including the central point-like object and the disk contribution \citep{bertin99}. 
We fit the two models to the rotation curve obtained in \cite{paneque20} (following the method proposed by \citealt{pinte18} to derive the height of the CO emitting layer) from the gas CO observations. 

\section{The rotation curve of a protoplanetary disk} \label{sec:style}

For a cool, slowly accreting disk, the centrifugal balance requires:
\begin{equation}
    \Omega^{2} = \frac{1}{R} \frac{\mathrm{d} \Phi_{\sigma}}{\mathrm{d} R}(R,z) +\frac{\mathcal{G} M_{\star}}{(R^{2}+z(R)^2)^{3/2}} + \frac{1}{R}\frac{1}{\rho}\frac{{\rm d}P}{{\rm d}R}
\label{eq:rotation}
\end{equation}
where $M_{\star}$ is the mass of the central object, $\Phi_{\sigma}$ is the disk contribution to the gravitational potential and where we also consider the pressure gradient (under the assumption of a barotropic disk). However, we expect the contribution of the pressure gradient to the rotation curve to be negligible when compared to the disk self-gravity contribution. Indeed, for a marginally stable self-gravitating disk the disk contribution is of the order of $H/R$, while the pressure term is $O(H^2/R^2)$, where $H$ is the pressure scale height \citep{kratter16}. To compute the pressure gradient we consider a disk temperature profile $T(R)\propto R^{-q}$, with $q=0.5$ (with $T=25$ K at $R=60$ au, corresponding to a disk aspect ratio at this location of $H/R=0.11$, \citealt{perezelias227}). %which corresponds to a sound speed $\sim 0.2-0.3$ km/s, and to a disk aspect ratio $H/R\sim0.15-0.11$ (computed at the inner and outer radius considered in this work $R\sim 60-300$ au).
%{\bf However, we neglect the pressure related term, since for a marginally stable self-gravitating disk the disk contribution is of the order of $H/R$, while the pressure term is $O(H^2/R^2)$, where $H$ is the pressure scale height \citep{kratter16}. 
We consider two models for the rotation curve of the disk orbiting around Elias 2-27: a Keplerian disk model and a self-gravitating disk model. Usually the Keplerian model is considered when $M_{\rm disk}\ll M_{\star}$, since in this case the contribution of the disk to the gravitational field is negligible \citep{pringle81}. The Keplerian model has also been used by \cite{paneque20} to estimate the stellar mass. 

In polar cylindrical coordinates, the radial gravitational field generated by the disk can be written as:
\begin{equation}
%\begin{split}
\frac{\partial \Phi_{\sigma}}{\partial R}(R, z) = \frac{\mathcal{G}}{R} \int_{0}^{\infty} \Bigg[ K(k)-\frac{1}{4}\left(\frac{k^{2}}{1-k^{2}}\right) \times  \left(\frac{R^{\prime}}{R}-\frac{R}{R^{\prime}}+\frac{z^{2}}{R R^{\prime}}\right) E(k)\Bigg] \sqrt{\frac{R^{\prime}}{R}} k \sigma \left(R^{\prime}\right) d R^{\prime}
\label{eq:sgpot}
%\end{split}
\end{equation}
%\begin{eqnarray}
%\frac{\partial \Phi_{\sigma}}{\partial R}(R, z) = %\frac{\mathcal{G}}{R} \int_{0}^{\infty} \Bigg[ %K(k)-\frac{1}{4}\left(\frac{k^{2}}{1-k^{2}}\right) %\times  \\ \left(\frac{R^{\prime}}{R}-\frac{R}{R^{\prim%e}}+\frac{z^{2}}{R R^{\prime}}\right) E(k)\Bigg] %\sqrt{\frac{R^{\prime}}{R}} k \sigma %\left(R^{\prime}\right) d R^{\prime} \nonumber
%\label{eq:sgpot}
%\end{eqnarray}
where $E(k)$ and $K(k)$ are complete elliptic integrals of the first kind, and $k^2 = 4RR^{\prime}/[(R + R^{\prime})^2 + z^2]$ (see \citealt{gradshteyn80}). \cite{ bertin99} computed the field ${\rm d}\Phi_{\sigma}/{\rm d}R$ in the equatorial plane by taking the limit $z \rightarrow 0$. Instead, we are interested in computing the rotation curve for the gas at a given height. The vertical position $z(R)$ has been determined by \cite{paneque20}, tracing the emitting layers of the CO-isotopologues channel maps with the method outlined in \cite{pinte18}, and has been parameterized as:
\begin{equation}
    z(R)=z_{0}\left(\frac{R}{R_{0}}\right)^{\psi}+z_{1}\left(\frac{R}{R_{0}}\right)^{\varphi}\,,
    \label{eq:z}
\end{equation}
where $z_0,z_1,\phi,\psi$ are fitting parameters reported in Table 2 of \cite{paneque20} and $R_0$ is equal to 115.88 au. Note that a major finding of \citet{paneque20} is that the West and East side of the disk show an asymmetry in the height of the gas layer, so the fitting parameters differ for the two sides of the disk. Furthermore, the two isotopologues considered ($^{13}$CO and C$^{18}$O) trace different vertical layers of the disk, and thus, will have distinct fitting parameters.
We also take into account the vertical position of the gas $z(R)$ when computing the Keplerian gravitational field in Eq.~\ref{eq:rotation}, as
\begin{equation}
\Omega_{\rm Kep}^2=\frac{\mathcal{G}M_{\star}}{(R^2+z(R)^2)^{3/2}}\,,
\label{eq:kepot}
\end{equation}
where $z(R)$ is defined in Eq.~\ref{eq:z}. The total disk surface density profile has been chosen after \cite{perezelias227} and \cite{andrews09} as, 
\begin{equation}
    \Sigma(R)=\Sigma_{c}\left(\frac{R}{R_{c}}\right)^{-p} \exp \left[-\left(\frac{R}{R_{c}}\right)^{2-p}\right]\,,
\label{eq:sigma}
\end{equation}
where $\Sigma_c$ is a normalisation constant assumed to be a free parameter of the model, while $R_c=200$ au is the truncation radius and the power-law index is fixed at $p=1$. We choose these values for the parameters to match the ones that were parametrized by \cite{perezelias227,paneque20}.% and then used in the hydrodynamical simulations performed by \cite{paneque20}. 

\section{Results} \label{sec:results}

Rather than performing a complete analysis of the channel maps, as done in \cite{paneque20}, we here take their constraints for the rotation curve and directly fit such rotation curve with two analytical competing models, the self-gravitating (see Eqs.~\ref{eq:rotation} and~\ref{eq:sgpot}) and the Keplerian one (see Eq.~\ref{eq:kepot}), using an MCMC algorithm as implemented in \textsc{emcee} \citep{emcee}. We choose 300 walkers and 3500 steps (where the convergence has already been reached at $\sim$2500 steps).  We also compare the data with a simple power-law fit, given by
\begin{equation}
    f(R)=p_1 \cdot R^{-p_2}\,
\label{eq:powerfit}
\end{equation}
where $p_1$ is a normalisation constant and $p2$ the power-law slope. An exponent $p_2=0.5$ would point to a Keplerian disk, while $p_2<0.5$ to the presence of a self-gravitating disk. Instead, an exponent $p_2>0.5$ could be suggesting a warp or the presence of chaotic accretion from the cloud (at large scale). In the analysis presented below, we do not fit the data points in the inner 60 au, since this region is strongly affected by the observed dust gap \citep{huang18b,paneque20}, and shows noisier data. However, we have also performed the fit also including this region, obtaining results similar to those showed in the following Sections. 

A detailed description of the velocity data used here can be found in \citet{paneque20}, along with the procedure used to obtain the height of the CO emitting layer, and we refer the reader to that paper for details. Here we just point out that, due to their higher signal-to-noise, the error bars are much lower for the $^{13}$CO data than for the C$^{18}$O data. Also note that we do not radially bin the data points obtained by \citet{paneque20} and that different velocity points related to the same radius arise from different azimuthal angles, highlighting the intrinsic non-axisymmetry of the disk. Still, our fitting model is by construction axisymmetric, since we are interested in the overall gravitational field of the disk, and we thus expect some non-negligible residuals to our fitting procedure due to this.

\begin{figure*}
	\includegraphics[scale=0.6]{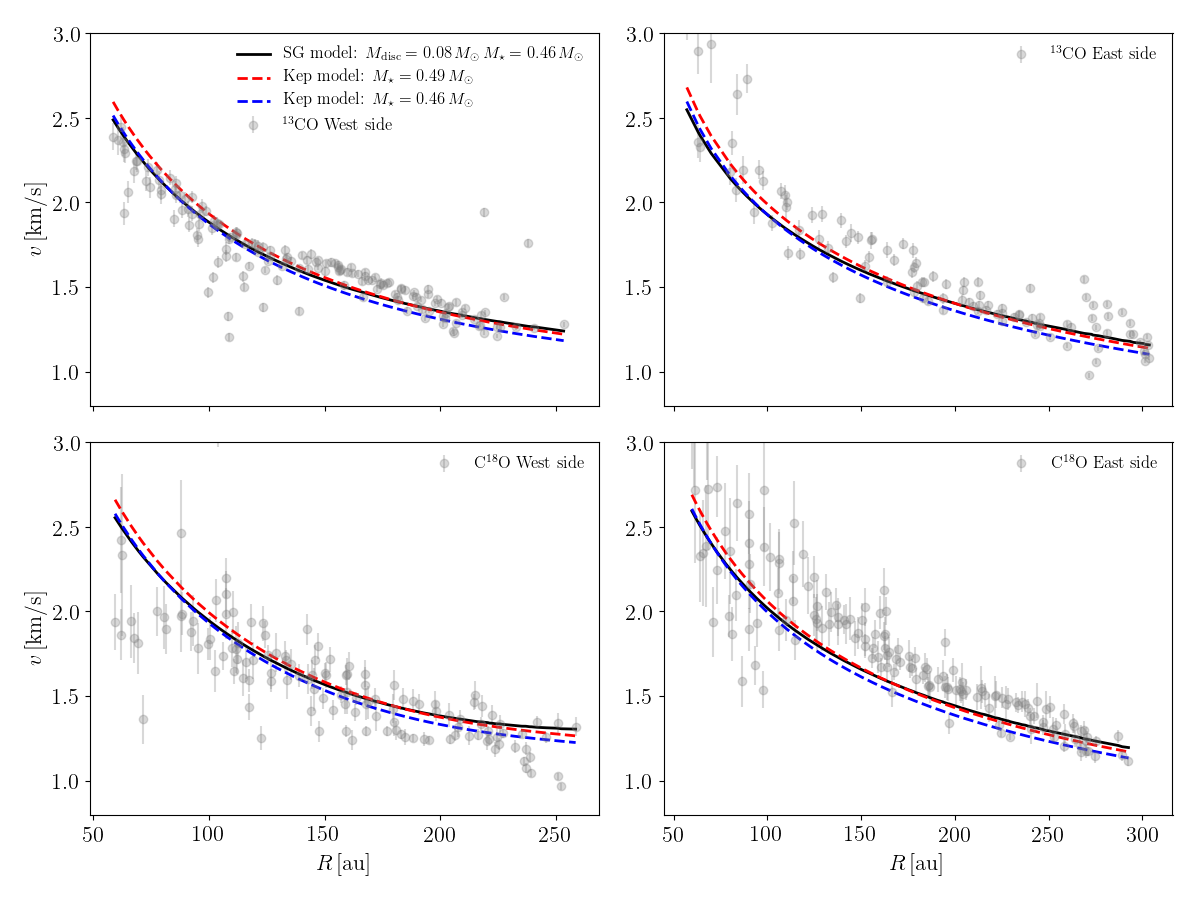}
 	\caption{Rotation curve for the $^{13}$CO and C$^{18}$O isotopologues for different models. The fitting procedure has been done simultaneously for the East (right column) and West (left column) side velocity data (plotted as grey markers), and for the two CO-isotopologues (top: $^{13}$CO; bottom: C$^{18}$O). The black solid line corresponds to the self-gravitating fit, the red dashed line to the Keplerian fit and the blue dashed one to the Keplerian curve obtained with the star mass from the SG fit. }
    \label{fig:all_fit}
\end{figure*}

\subsection{Combined fit}\label{subsec:all}

In Fig.~\ref{fig:all_fit} we show the results obtained with the Keplerian (see Eq.~\ref{eq:kepot}, red and blue dashed lines) and self-gravitating (see Eqs.~\ref{eq:rotation} and~\ref{eq:sgpot}, black solid lines) models, when simultaneously fitting all the data points (but considering the height profiles separately) for both CO isotopologues and for both sides of the disk, shown in separate panels for clarity, with the East and West side of the disk on the right and left columns, respectively, and the two CO-isotopologues (top: $^{13}$CO; bottom: C$^{18}$O). 
The red line corresponds to the Keplerian best fit model, the blue line shows the rotation curve for a Keplerian disk where the star mass has been fixed to the one found with the self-gravitating best fit. The values obtained for the fit parameters are reported in the third column of Table \ref{tab:kepmodel}. The East data points, especially for the C$^{18}$O, tend to lie above the best fit curves, since in this combined fit the model naturally tends to reproduce the lower uncertainty $^{13}$CO data. If we first look at the power-law model, once we leave the freedom of a general power-law index, the best-fit value of the exponent $p_2$ is smaller than $0.5$ ($p_2 = 0.43\pm 0.03$), by more than $2\sigma$. This already suggests that the data are better reproduced by a self-gravitating model. In such a model, the disk mass obtained from the combined fit is $M_{\rm disk}=0.08\pm 0.04\,M_{\odot}$ with a star mass $M_{\star}=0.46\pm 0.03\,M_{\odot}$. Note that we obtain a non-zero measurement of the disk mass too within $\sim 2\sigma$ uncertainties.
Instead, in the Keplerian case, the star mass is $M_{\star}=0.49\pm0.01\,M_{\odot}$. For both models, the stellar mass  is in agreement with previous estimates \citep{paneque20}. 

\begin{deluxetable}{cccc}
\label{tab:kepmodel}
%\tablenum{1}
\tablecaption{Parameter obtained with a Keplerian, self-gravitating and power-law model for each CO-isotopologues (first two columns) and for a combined fit (both sides and both CO-isotopologues, third column). We also show the reduced $\chi^2$ difference between the Keplerian and self-gravitating fit, as $\lambda = \Delta (\chi^2_{\rm red})$.}
\tablewidth{0pt}
\tablehead{
 & \colhead{$^{13}\mathrm{CO}$} & \colhead{$\mathrm{C}^{18}\mathrm{O}$} & \colhead{Combined fit} \\
% \hline
 %\colhead{\textbf{Keplerian fit}} &   &  & \\
 }
\startdata
\textbf{Keplerian fit} &   &  &  \\
\hline
$M_{\star}\,[M_{\odot}]$ & 0.50$^{+0.01}_{-0.01}$ & 0.46$^{+0.03}_{-0.03}$ & 0.49$^{+0.01}_{-0.01}$ \\
 \hline
  \hline
 \textbf{Self-gravitating fit} &   &  &  \\
 \hline
    $M_{\star}\,[M_{\odot}]$ & $0.45^{+0.03}_{-0.03}$ & $0.43^{+0.05}_{-0.07}$ & $0.46^{+0.03}_{-0.03}$ \\
    $M_{\rm disk}\,[M_{\odot}]$ & $0.1^{+0.05}_{-0.04}$ & $0.08^{+0.08}_{-0.05}$ & $0.08^{+0.04}_{-0.04}$ \\
 \hline
  \hline
 \textbf{$\lambda = \Delta$($\chi^2_{\rm red}$)} & 2.16   & -0.19 & 1.38  \\
  \hline
  \hline
\textbf{Power-law fit} &   &  & \\
\hline
$p_1$  & $13.46^{+2.39}_{-2.07}$ & $25.31^{+14.67}_{-9.39}$ & $13.95^{+2.39}_{-2.06}$ \\
$p_2$ & $0.43^{+0.03}_{-0.03}$ & $0.54^{+0.09}_{-0.1}$ &  $0.43^{+0.03}_{-0.03}$ \\
 \hline
 \enddata
\end{deluxetable}

\begin{figure*}
	\includegraphics[scale=0.6]{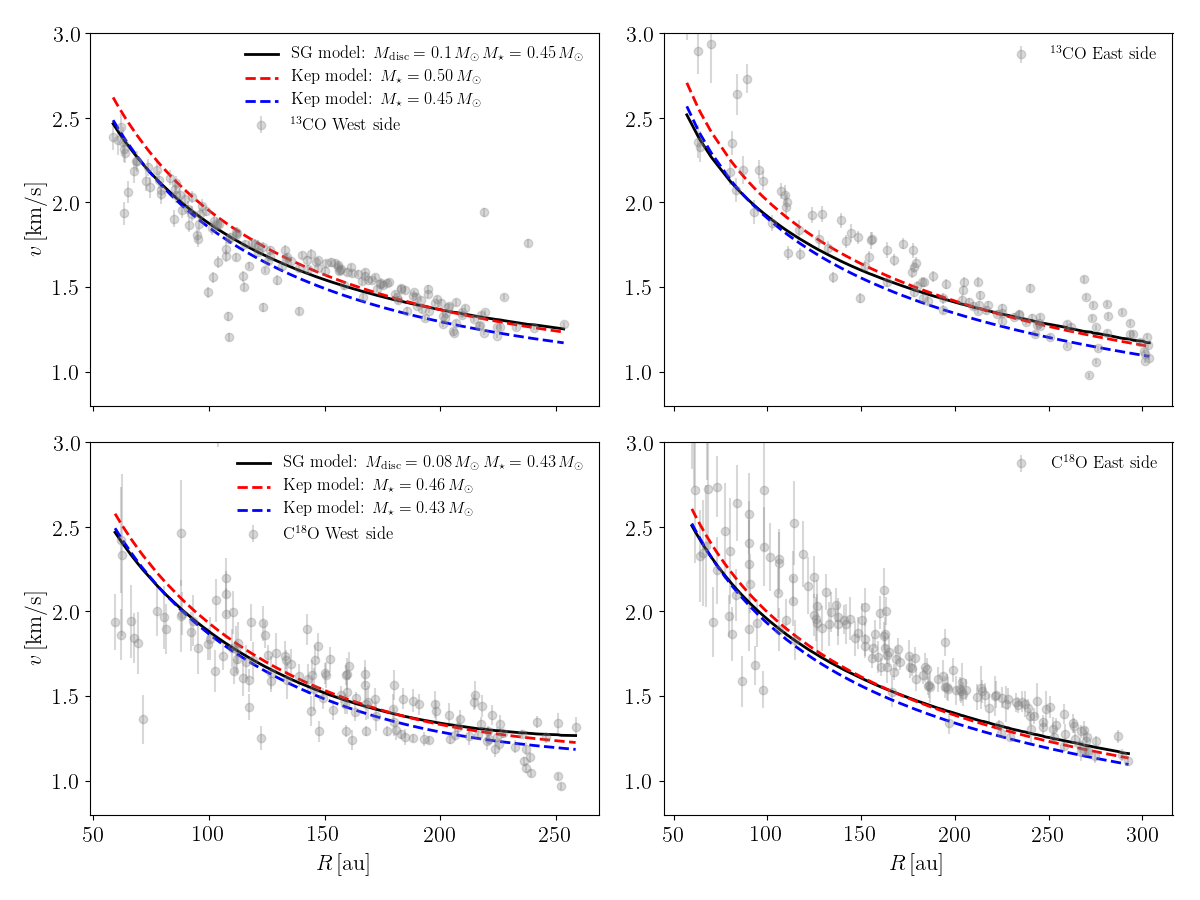}
 	\caption{Rotation curve for the $^{13}$CO and C$^{18}$O isotopologues for different models. The fitting procedure has been done simultaneously for the East (right column) and West (left column) side velocity data (plotted as grey markers). The black solid line corresponds to the self-gravitating fit, the red dashed line to the Keplerian fit and the blue dashed one to the Keplerian curve obtained with the star mass from the SG fit.
 	}
    \label{fig:both_sides}
\end{figure*}

\subsection{Individual isotopologues fit}\label{subsec:res_both}
We also performed a fit separately for the two CO isotopologues. The result is shown in Fig.~\ref{fig:both_sides}, where the upper and lower panels correspond to the West (left panel) and East (right panel) side of the $^{13}$CO and  C$^{18}$O isotopologue, respectively.
The parameters obtained from the best-fit models are shown in the first two columns of Table~\ref{tab:kepmodel}. Also in this case, we start by looking at the power-law model. For the $^{13}$CO data, the best-fit value of the exponent $p_2$ is again smaller than $0.5$ ($p_2 = 0.43\pm0.03$), meaning that the self-gravitating model should be preferred to reproduce the data. In contrast, for the C$^{18}$O the best-fit power-law index is $p_2=0.54^{+0.09}_{-0.1}$, where the value of 0.5 is inside the uncertainties, meaning that a purely Keplerian model is consistent with the $^{18}$CO data, given the larger uncertainty in the velocity points in this case. We also note that the obtained value $>0.5$ could suggest the presence of a warp or chaotic accretion from the cloud. By considering the results for the $^{13}$CO, best fitted by a self-gravitating model, the disk mass is $M_{\rm disk} = 0.1^{+0.05}_{-0.04}\,M_{\odot}$, with a stellar mass of $M_{\star}=0.45\pm0.03\,M_{\odot}$. 
\section{Discussion} \label{sec:discussion}

Having performed fits for the self-gravitating and the Keplerian model, we now compare which one is a better fit to the data%compare the goodness of each one
. To do so, we compute the reduced $\chi$-square ($\chi^2_{\rm red}$) for each model and each CO-isotopologue. We then compute  the likelihood ratio $\lambda$, defined as the difference between the Keplerian and self-gravitating minimum reduced $\chi$-square:
\begin{equation}
    \lambda = (\chi^2_{\rm red})_{\rm min, Kep} - (\chi^2_{\rm red})_{\rm min, SG}\,.
\end{equation}
For Gaussian, independent measurements, this function is distributed like a $\chi^2$ with $n$ degrees of freedom, where $n$ is the number of new parameters in the more general case ($n=1$), with the hypothesis that the less general model is correct (i.e. the Keplerian model). 
The computed values are presented in Table~\ref{tab:kepmodel}. We obtain $\lambda\simeq 2.16$ for the $^{13}$CO fit, and $1.38$ for the combined fit, which means that the Keplerian model is rejected with respect to the self-gravitating one. If we consider only the $\mathrm{C}^{18}\mathrm{O}$ data, instead, the likelihood ratio tends to slightly prefer a simple Keplerian model (see Table~\ref{tab:kepmodel}). This means that in this case the two models are indistinguishable, possibly because the errors are larger with respect to the $^{13}\mathrm{CO}$ case. 

In summary, the best fitting model for the combined set of data, including both available CO isotopologues is a non-Keplerian one, with a 
disk mass $M_{\rm disk}=0.08\pm0.04\,M_{\odot}$, and a star mass $M_{\star}=0.46\pm0.03\,M_{\odot}$. Considering only the $^{13}$CO data (that are of better quality with respect to the C$^{18}$O ones), we obtain a disk mass of $M_{\rm disk}=0.1^{+0.05}_{-0.04}\,M_{\odot}$ and a star mass $M_{\star}=0.45\pm0.03\,M_{\odot}$. In both cases, we obtain a non-zero disk mass within $2\sigma$. Instead, the C$^{18}$O alone might be compatible with a purely Keplerian rotation curve, even though the self-gravitating fit returns a non-zero disk mass to within 1$\sigma$ uncertainty, $M_{\rm disk}= 0.08^{+0.08}_{-0.05}\,M_{\odot}$ and a star mass of $0.43^{+0.05}_{-0.07}\,M_{\odot}$. 

Thus, assuming a total disk mass equal to $0.08-0.1$ $M_{\odot}$ (and a star mass of $0.46-0.45$ $M_{\odot}$) as obtained from the fits above, we get a disk-to-star mass ratio of $\sim 0.17-0.22$. Gravitational instabilities arise when the disk-to-star mass ratio becomes of the order of the disk aspect ratio $H/R$, which is typically of the order of $\approx 0.1$ for protostellar disks. The disk mass we derive from the rotation curve is thus in the correct range to produce gravitational instabilities and thus the spiral structure observed. In particular, the observed two-armed grand-design structure is strongly suggestive of an internal origin due to gravitational instabilities. We note that from the relation between the disk-to-star mass ratio and the number of spiral arms $M_{\rm disk}/M_\star\propto 1/m$ \citep{lodato04,CLC09,dong15} the obtained disk mass would point to high $m$ modes, while just two spiral arms are observed through ALMA. However, \cite{dipierro14} have demonstrated that, even if the density structure has an intrinsic $m>2$ spiral, smaller-scale arms can be washed out by the limited resolution of the instrument, leaving only the lowest $m$ modes in ALMA dust continuum observations. The disk mass obtained in this work is consistent with those used in the hydrodynamical simulations that reproduce the observed spirals, as performed by \cite{paneque20}, where they employ a disk-to-star mass ratio in the range of q = 0.1-0.3, and in the simulations of \cite{cadman20}, with a slightly larger q = 0.27 value.
 
Having obtained a dynamical estimate of the total disk mass, and assuming a dust disk mass of $10^{-3}\,M_{\odot}$ \citep{paneque20,perezelias227}, we can put interesting constraints on the gas-to-dust, that turns out to be of the order of $\approx 80-100$ (for the combined fit and the $^{13}$CO isotopologue), which in the first case corresponds to a factor $\sim1.2$ smaller than the usually assumed value of 100. Note that the so obtained gas-to-dust ratio estimate extremely depends on the dust mass derivation and thus it should be considered with care. For this derivation we assumed a dust mass of $10^{-3}\,M_{\odot}$, but \cite{paneque20} showed that the disk being optically thick with a low spectral index, scattering could be important leading to a dust mass estimate up to 2 times larger than previously considered. 

\begin{figure*}[ht!]
\centering
	\includegraphics[scale=0.55]{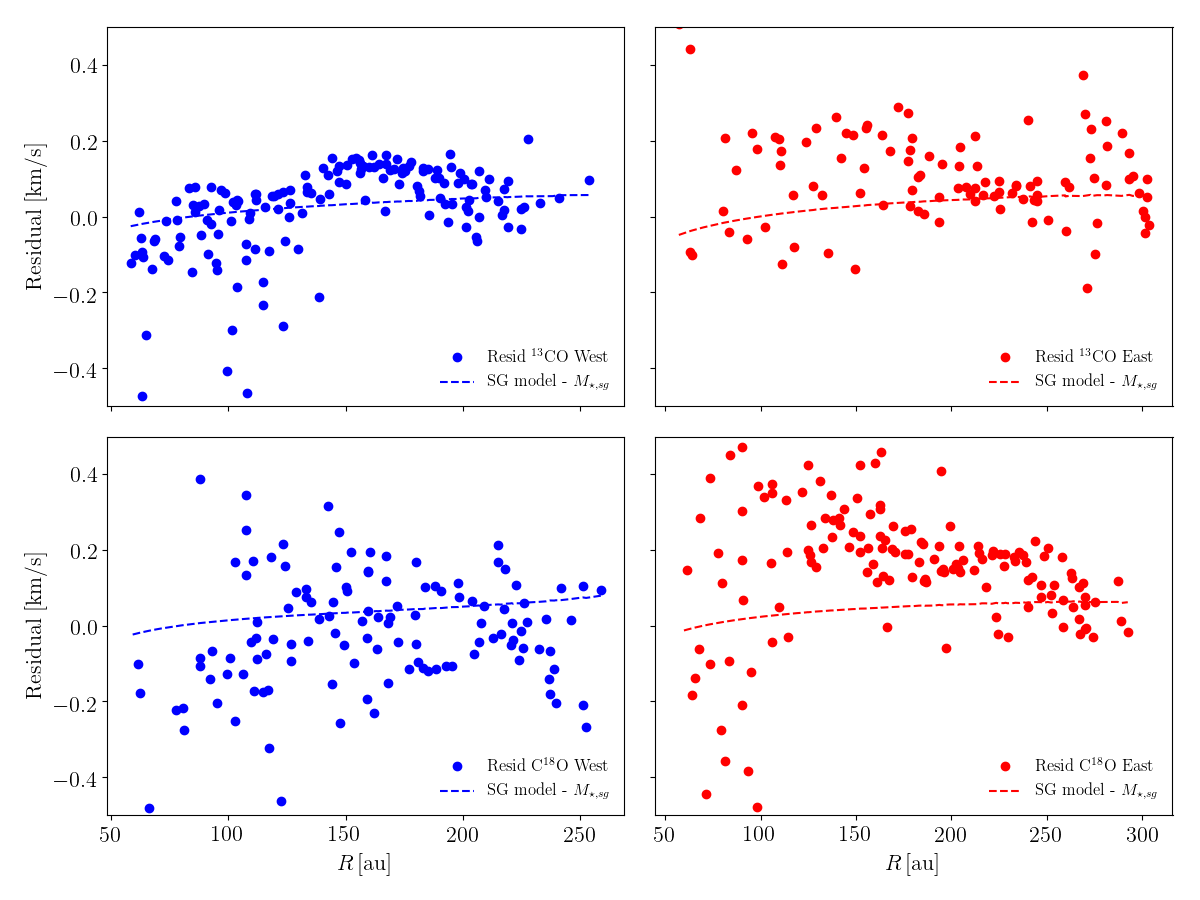}
 	\caption{Residuals obtained for the combined fit. The top panels show residual for the $^{13}$CO (blue points and dashed line), while the bottom ones for the C$^{18}$O (red points and dashed line). The left panels correspond to the West side, while the right one to the East side. Points represent the difference between the velocity data and a Keplerian model where the star mass $M_{\star,{\rm sg}}$ has been obtained through the self-gravitating model. The dashed line is the difference between the self-gravitating model and the above mentioned stellar contribution, $M_{\star,{\rm sg}}$.
 	}
    \label{fig:residual_all}
\end{figure*}
As a further analysis of the results obtained in the combined fit, we show in Fig.~\ref{fig:residual_all} the residuals for both disk sides (left: West; right: East) and the CO-isotopologues (blue: $^{13}$CO; red: C$^{18}$O). In particular, the points are the difference between the velocity data and a Keplerian model where the star mass $M_{\star,{\rm sg}}$ has been obtained through the self-gravitating model. The dashed line is the difference between the self-gravitating model and the Keplerian velocity with the stellar mass $M_{\star,{\rm sg}}$. From these results it appears that especially the residuals for the West side in the $^{13}$CO do require a significant disk contribution to the gravitational potential. Indeed, the data residuals present an increasing trend, following the disk contribution model, in particular for the West side. Instead, for the C$^{18}$O there is still some large scatter in both directions. This aspect is particularly interesting, indeed, \cite{paneque20} find that there is an important asymmetry between the East and West side data (see their Discussion section). The main characteristic of this asymmetry is that the West side is more compact and brighter than the East side, which is more extended and cloud-contaminated. The rotation curve on the East side then should be considered with care, since the disk can be contaminated by chaotic accretion from the cloud. This infall of material could in principle change the centrifugal balance, increasing the complexity of the system. 
For this reason, we decided to repeat the fit procedure for the West side only. The obtained results are described in Appendix~\ref{sec:singleside}. We obtain an even stronger indication in favor of a self-gravitating fit, with the Keplerian fit rejected with 80\% confidence for the combined fit and with 97\% confidence considering the $^{13}$CO data only. The resulting disk mass in this case is $M_{\rm disk}=0.16\pm0.06\,M_{\odot}$ with a stellar mass $M_{\star}=0.41\pm0.04\,M_{\odot}$, and thus in a disk-to-star mass ratio of $\sim0.40$. 

Finally, it has to be noted that small scale gas turbulence could contribute to deviations from keplerian motion, but this generally amounts to no more than $\ 0.1 c_{\rm s} \sim 20$ m/s \citep{flaherty20}, being thus smaller than the observed deviation (the disk contribution is $\sim 50-100$ m/s, see Fig.~\ref{fig:residual_all}). %We highlight that we also per  the pressure gradients contribution might have some impact in our estimate of the disk mass. However, if anything, such contribution would further enhance our disk mass estimate, since while the disk self-gravity brings a super-Keplerian contribution to the rotation curve, the pressure gradients contribution is sub-Keplerian. We will address this issue in a future work.

\section{Conclusions} 
In this paper we have looked for deviations from Keplerian rotation in the disk orbiting around the Elias 2-27 system, providing for the first time a dynamical measurement of the total mass of a planet forming disk, by fitting its rotation curve as derived from CO emission with a model including both the stellar and the disk contribution to the gravitational field. 
We performed three different fit procedures, that is, a combined fit considering both disk sides and both CO-isotopologues, an individual fit to the data points for the separate isotopologues, and a third fit considering only the less cloud-contaminated West side of the disk; the last case is described in Appendix C. 
The outcome of these analysis is that the $^{13}$CO isotopologues data, and in particular the West side of the disk, are better reproduced by a self-gravitating disk model rather than a pure Keplerian one. The same is true also considering both isotopologues, although with smaller confidence, with a resulting disk mass of $M_{\rm disk}=0.08\pm0.04\,M_{\odot}$ and a stellar mass $M_{\star}=0.46\pm0.03\,M_{\odot}$ (where the stellar mass is compatible with previous estimates). We point out that we obtain a non-zero measure of the disk mass within 2$\sigma$ uncertainty%(CL of 95\%)
, both in the combined fit and in the fit for the $^{13}$CO isotopologue alone. Assuming these values for the disk and star mass, and assuming a dust disk mass of $10^{-3}\,M_{\odot}$ \citep{paneque20,perezelias227}, we obtain a disk-to-star mass ratio of $\simeq$ 0.17 and a gas-to-dust ratio of $\simeq 80$. These results highlight the fact that Elias 2-27 should be considered as a self-gravitating disk, reinforcing the internal gravitational instability interpretation for the observed spiral structures. This result is more evident when fitting for the $^{13}$CO data on the West side of the disk, that are less contaminated by the cloud contamination and possible infall \citep{paneque20} for which we obtain a disk-to-star mass ratio of $\sim 0.40$, with a disk mass of $0.16\,M_{\odot}$ and a star mass of $0.41\,M_{\odot}$. We point out that the lower confidence level obtained in the combined isotopologues fit is due to the relatively lower quality of the C$^{18}$O data, which can be improved in future observations. 

Finally, we remark that this method to estimate the disk mass can be applied to other protoplanetary disks (such as, for example, IM Lup and WaOph 6 from the DSHARP sample, \citealt{huang18c}, and RU Lup, \citealt{huang20}, that also show a prominent spiral structure), aiming to give better constraints (independent of CO or dust to H$_2$ conversion) on the disk mass. Such measurement can also be used to calibrate the conversion factors between dust and total mass, at least for these systems. 

\acknowledgments
The authors want to thank the referee for constructive comments that improved this manuscript. 
This paper makes use of the following ALMA data: \#2013.1.00498.S, \#2016.1.00606.S and  \#2017.1.00069.S. 
ALMA is a partnership of ESO (representing its member states), NSF (USA), and NINS (Japan), together with NRC (Canada),  NSC and ASIAA (Taiwan), and KASI (Republic of Korea), in cooperation with the Republic of Chile. 
The Joint ALMA Observatory is operated by ESO, AUI/NRAO, and NAOJ. 
This work has received funding from the European Union’s Horizon 2020 research and innovation programme under the Marie Skłodowska-Curie grant agreement No 823823 (RISE DUSTBUSTERS project). 
L.M.P.\ acknowledges support from ANID project Basal AFB-170002 and from ANID FONDECYT Iniciaci\'on project \#11181068.
We used the \textsc{emcee} algorithm \citep{emcee}, the \textsc{corner} package \cite{corner} to produce corner plots and  \textsc{python}-based MATPLOTLIB package \citep{matplotlib} for all the other figures.

\appendix

\section{Corner plot results for the combined fit}

\begin{figure*}[ht!]
\centering
	\includegraphics[scale=0.3]{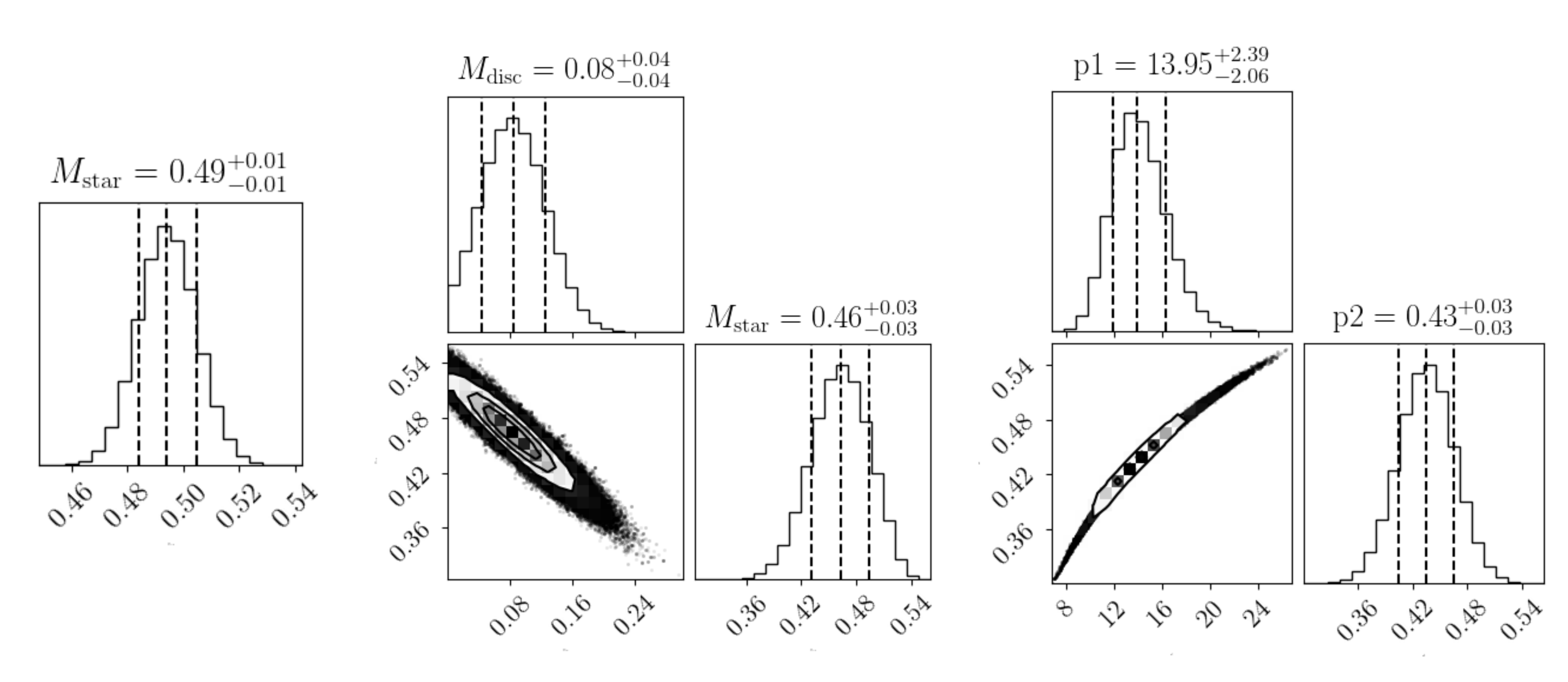}
 	\caption{Probability distribution and corner plots obtained with the Keplerian (left), disk+star (centre) and power law (right) model for the total fit, where the East and West sides, and the $^{13}$CO and C$^{18}$O data have been fitted simultaneously. 
 	}
    \label{fig:cornerplots_all}
\end{figure*}

Fig.~\ref{fig:cornerplots_all} collects the corner plots of the MCMC procedure for the Keplerian (left), self-gravitating (disk+star, centre) and power-law (right) fit of the CO isotopologues velocity data for the combined fit. The fit results correspond to those discussed in Sec.~\ref{subsec:all}. The final masses and errors are computed from the median value, 16th and 84th percentile uncertainties derived from the posteriors.

\section{Corner plot results for the individual isotopologues fit}
\begin{figure*}[ht!]
\centering
	\includegraphics[scale=0.3]{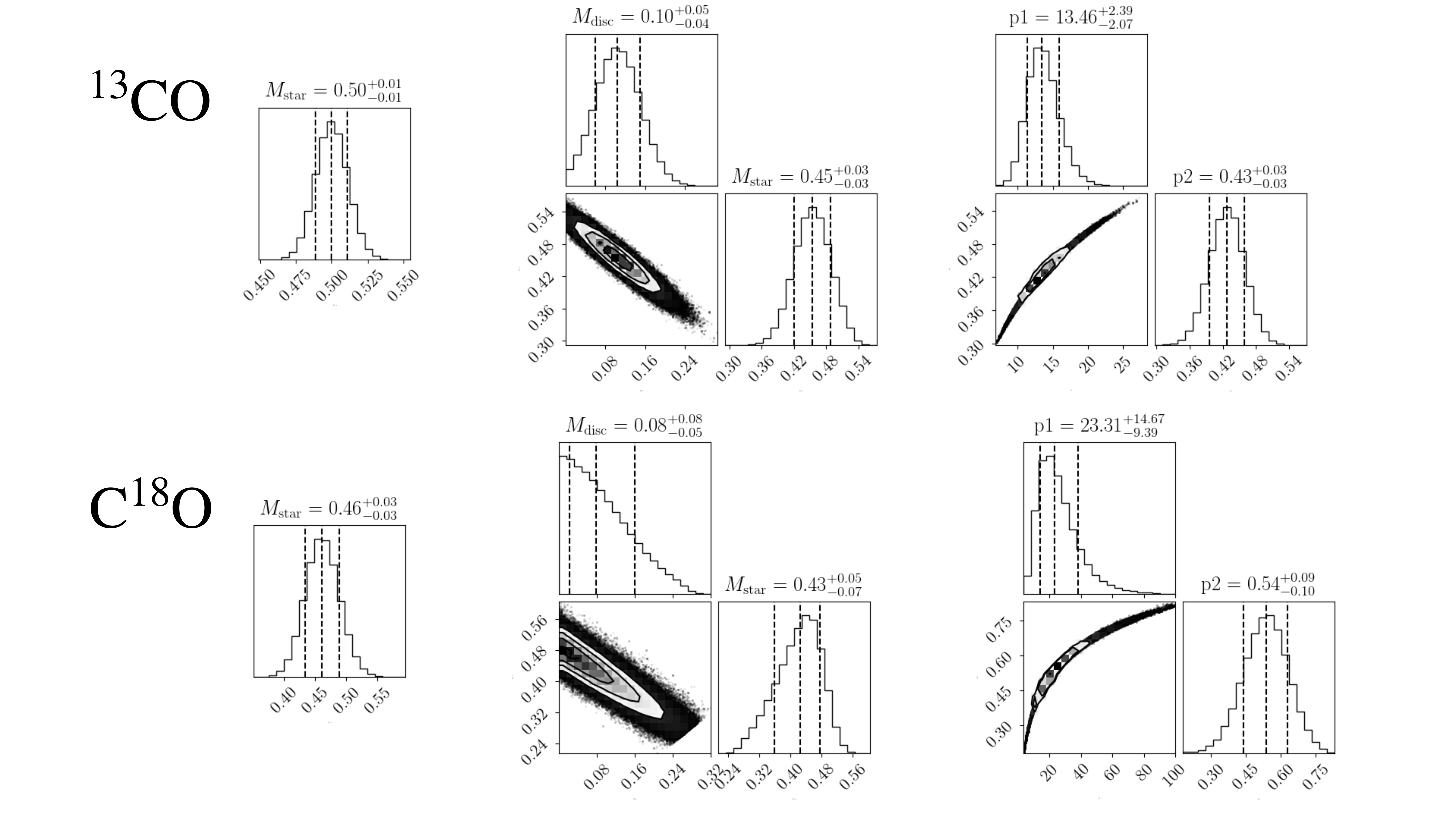}
 	\caption{Probability distribution and corner plots for different models (from left to right: Keplerian, disk+star,power law model) for the $^{13}$CO (upper panels) and  C$^{18}$O (lower panels) isotopologues.
 	}
    \label{fig:corner_bothsides}
\end{figure*}

Fig.~\ref{fig:corner_bothsides} collects the corner plots of the MCMC procedure for the Keplerian (left panels), self-gravitating (disk+star, central panels) and power-law (right panels) fit of the CO isotopologues velocity data (upper panels: $^{13}$CO; lower panels: C$^{18}$O), fitted separately. The fit results correspond to those discussed in Sec.~\ref{subsec:res_both}. The final masses and errors are computed from the median value, 16th and 84th percentile uncertainties derived from the posteriors. 

\section{West side fit\label{sec:singleside}}
\begin{figure*}[ht!]
	\includegraphics[scale=0.5]{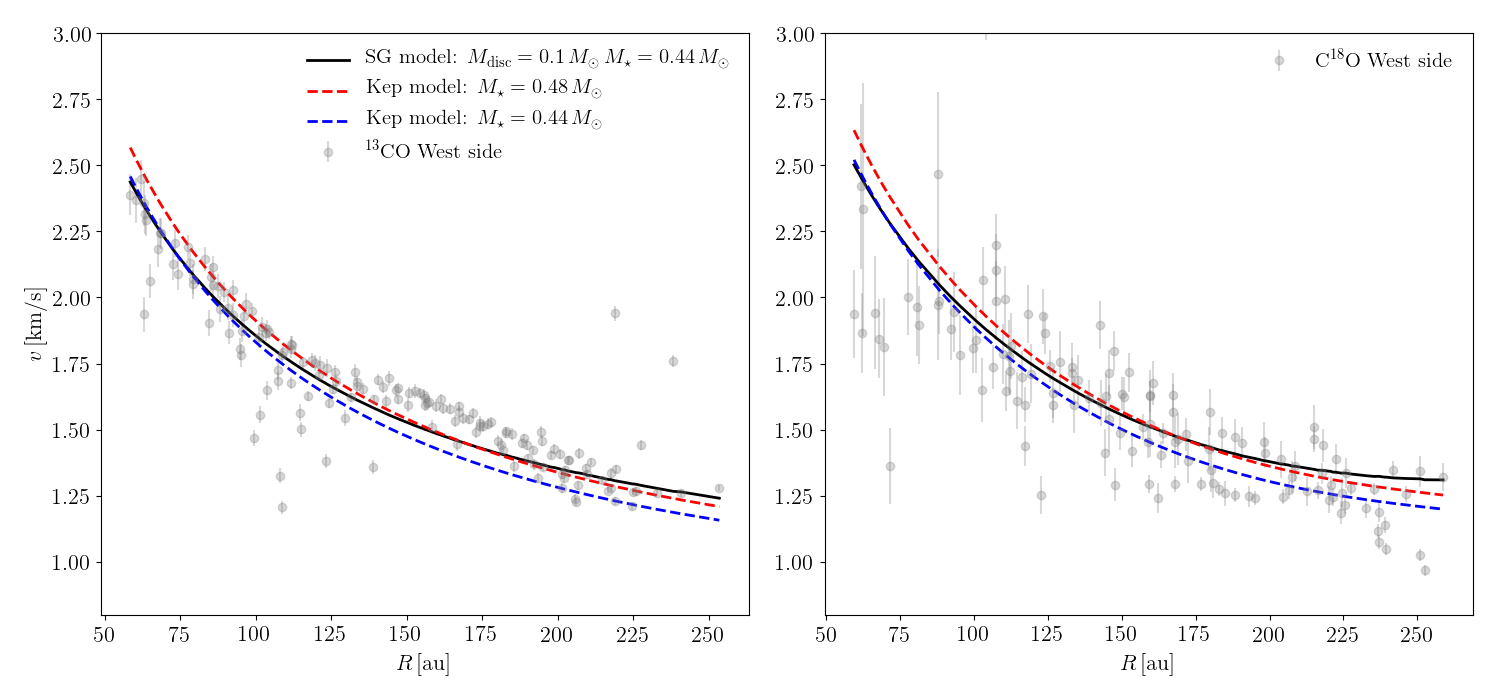}
	\includegraphics[scale=0.5]{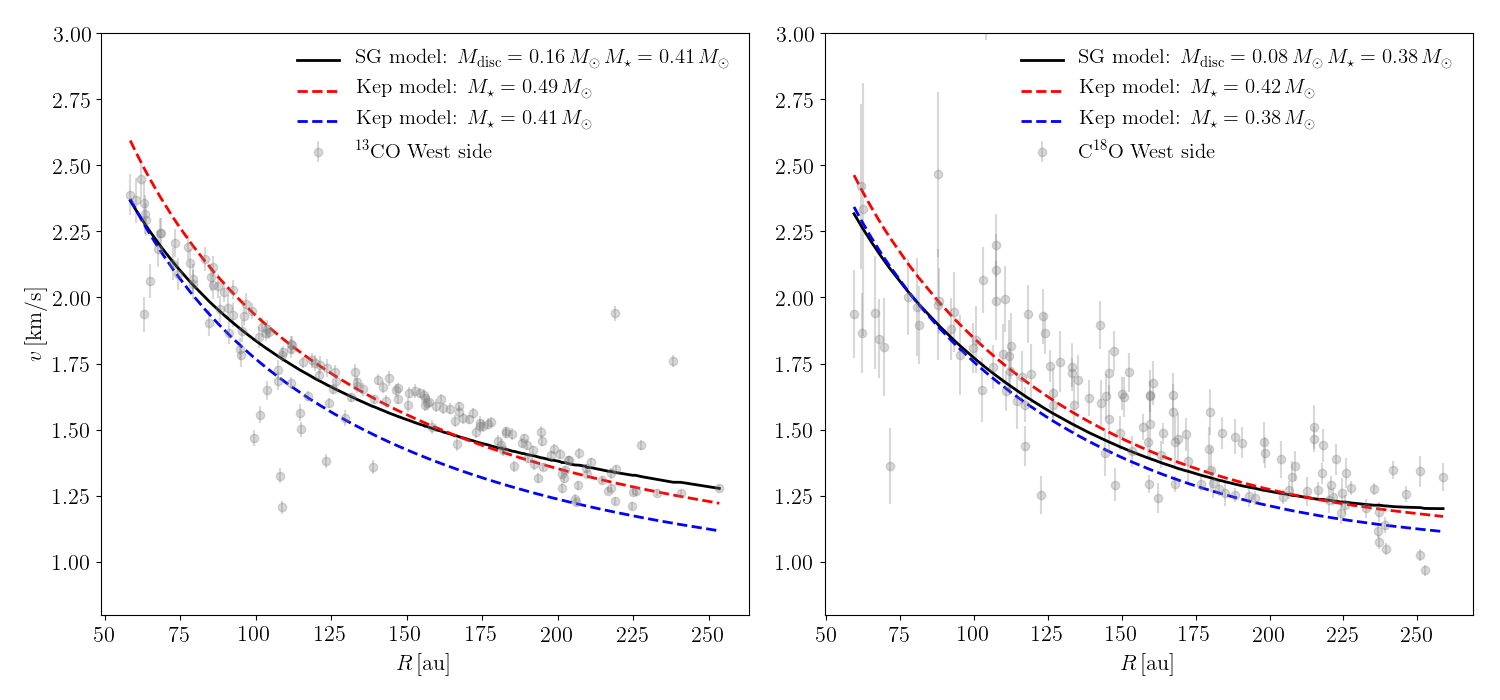}
 	\caption{Top panels: rotation curve for the $^{13}$CO (left) and C$^{18}$O (right) isotopologues for different models. In this case the fitting procedure has been performed considering only the West side of the disk. The black solid line corresponds to the self-gravitating fit and the red and blue dashed lines to the Keplerian fit and to the Keplerian curve obtained with the star mass from the SG fit, respectively. Grey markers represent the velocity data. Bottom panels: same as above, but with a fit performed separately for the two CO-isotopologues.
 	}
    \label{fig:disk_star_single}
\end{figure*}
In Fig.~\ref{fig:disk_star_single} we show results where the fit has been done only for the West side, which is less contaminated by the infalling cloud. As in Sec.~\ref{sec:results}, the models we choose are a self-gravitating model (black solid lines), a Keplerian one (red dashed line), and a simple power-law model (see Eq.~\ref{eq:powerfit}). As before, the blue dashed line represent the Keplerian rotation curve with the star mass fixed to the value obtained with the self-gravitating fit. The top panels collect results obtained for the combined fit, while the bottom panels show results when fitting separately the two CO-isotopologues. In Table~\ref{tab:kepmodel_sep} we report the obtained parameter values for each model, each CO-isotopologues (first two columns), and for the combined fit (fitting simultaneously the CO-isotopologues data, third column).

\begin{deluxetable*}{cccc}
 \label{tab:kepmodel_sep}
%\tablenum{2}
\tablecaption{System parameters obtained with a Keplerian, self-gravitating and power-law models for the West side disk fit. The first two columns collect results for a fit when considering separately the two CO-isotopologues, the third one when considering them simultaneously. We also show the reduced $\chi^2$ difference between the Keplerian and self-gravitating fit, as $\lambda = \Delta$($\chi^2_{\rm red}$).}
\tablewidth{0pt}
\tablehead{
 & \colhead{$^{13}\mathrm{CO}$} & \colhead{$\mathrm{C}^{18}\mathrm{O}$} & \colhead{Combined fit} \\
% \hline
 %\colhead{\textbf{Keplerian fit}} &   &  & \\
 }
\startdata
\textbf{Keplerian fit} &   &  &  \\
\hline
$M_{\star}\,[M_{\odot}]$ & 0.49$^{0.01}_{-0.01}$ & 0.42$^{0.03}_{-0.03}$ &  0.48$^{+0.01}_{-0.01}$  \\
 \hline
  \hline
 \textbf{Self-gravitating fit} &   &  &  \\
 \hline
   $M_{\star}\,[M_{\odot}]$ & $0.41^{+0.04}_{-0.04}$ & $0.38^{+0.06}_{-0.07}$ & $0.44^{+0.04}_{-0.04}$ \\
    $M_{\rm disk}\,[M_{\odot}]$ & $0.16^{+0.06}_{-0.06}$ &  $0.08^{+0.08}_{-0.06}$ &  $0.1^{+0.06}_{-0.05}$ \\
 \hline
  \hline
 \textbf{$\lambda = \Delta$($\chi^2_{\rm red}$)} & 4.57   &  -0.51 &  1.58  \\
  \hline
  \hline
\textbf{Power-law fit} &   &  & \\
\hline
  $p_1$  & $12.66^{+3.47}_{-2.75}$ & $27.49^{+23.67}_{-13.33}$ &  $14.07^{+3.53}_{-2.85}$ \\
    $p_2$ & $0.42^{+0.05}_{-0.05}$ & $0.58^{+0.12}_{-0.13}$ &  $0.44^{+0.04}_{-0.04}$  \\
 \hline
 \enddata
\end{deluxetable*}

The resulting disk mass in the combined fit is $M_{\rm disk}=0.1^{+0.06}_{-0.05}\,M_{\odot}$, with a stellar mass of $M_{\star}=0.44\pm0.04\,M_{\odot}$. The obtained likelihood ratio $\lambda = \Delta \chi^2_{\rm red} \sim 1.6$ implies a rejection of the Keplerian model.% with 80$\%$ confidence. 
This values would provide a disk-to-star mass ratio of 0.22, meaning that the disk mass is in the right range to give rise to gravitational instabilities. Also when considering the CO-isotopologues separately, this behaviour is seen for the $^{13}$CO. In this case, the expected disk mass is $M_{\rm disk}=0.16\pm0.06\,M_{\odot}$, the stellar mass $M_{\star}=0.41\pm0.04\,M_{\odot}$ and $\lambda = \Delta \chi^2_{\rm red}  \sim 4.6$, implying also in this case a rejection of the Keplerian model. %the Keplerian model is rejected with 97\% confidence. 
These values correspond to a disk-to-star mass ratio of 0.39. 
\begin{figure*}[h!]
\centering
	\includegraphics[scale=0.7]{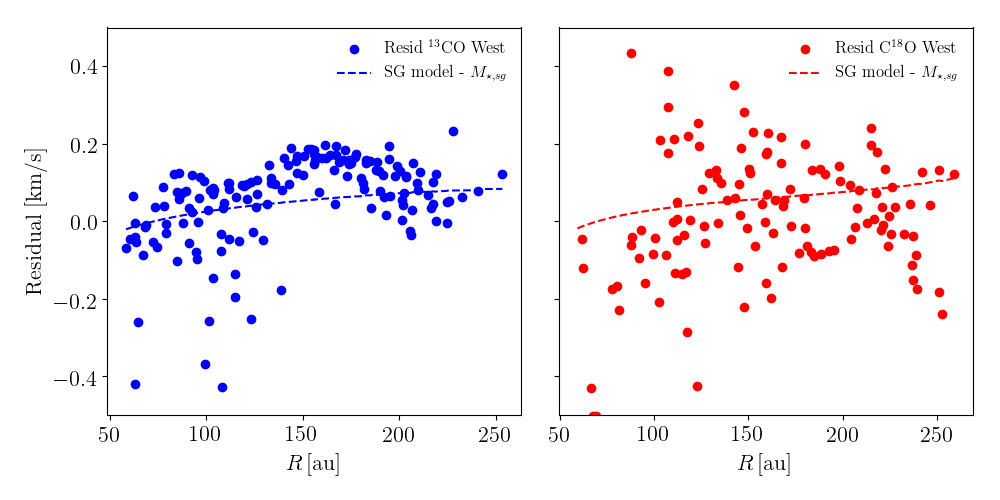}
 	\caption{Residuals obtained for the West side fit, when considering simultaneously the two CO-isotopologues. The left panel shows residual for the $^{13}$CO (blue markers and dashed line), while the right one for the C$^{18}$O (red markers and dashed line). Markers represent the difference between the velocity data and a Keplerian model where the star mass $M_{\star,{\rm sg}}$ has been obtained through the self-gravitating model. The dashed line is the difference between the self-gravitating model and the above mentioned star contribution, $M_{\star,{\rm sg}}$.
 	}
    \label{fig:residual_west}
\end{figure*}

For the combined (both CO- isotopologues) West fit, we collect in Fig.~\ref{fig:residual_west} the residuals. The difference between the self-gravitating model and the star contribution to the gravitational potential (as obtained with the self-gravitating model) shows that the disk contribution fits well the data residual (data - Keplerian mass), especially for the $^{13}$CO isotopologue.

\subsection{Corner plot results}
\begin{figure*}[ht!]
\centering
	\includegraphics[scale=0.3]{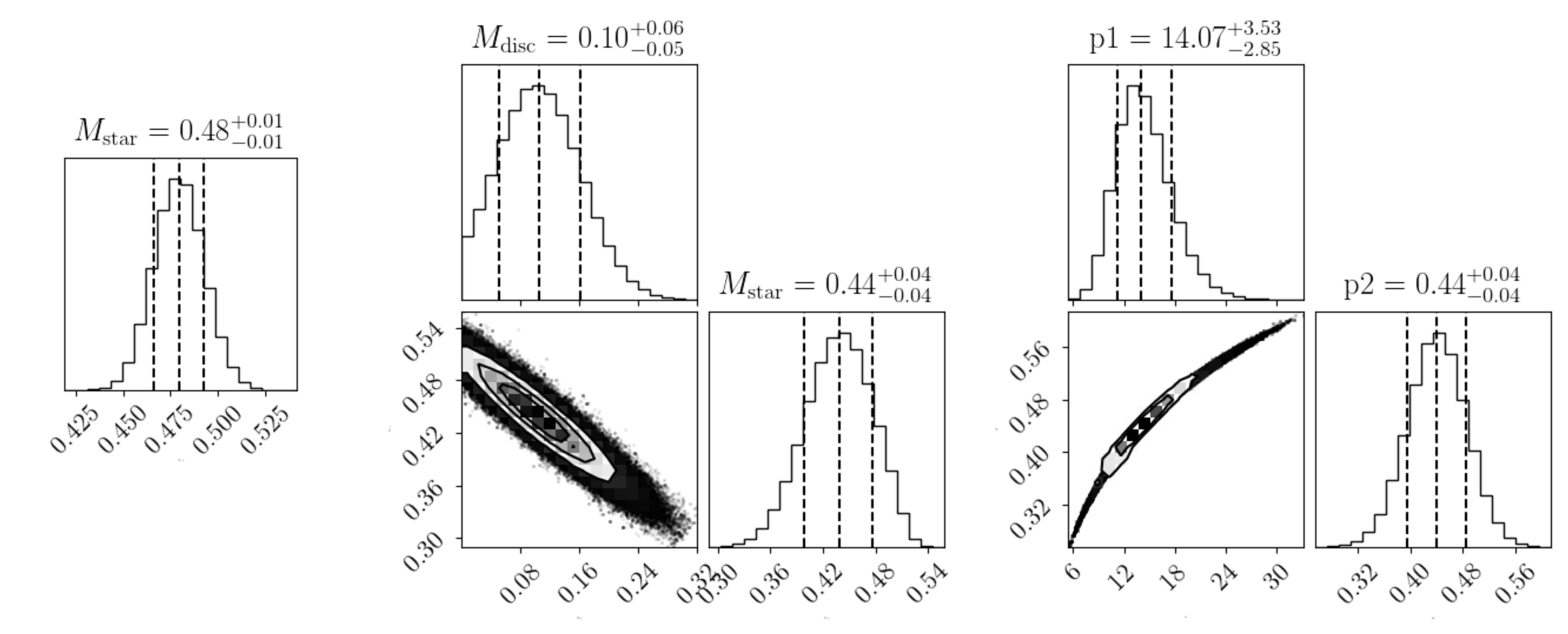}
 	\caption{Probability distribution and corner plots obtained with the Keplerian (left), disk+star (centre) and power law (right) model for the West side, where the $^{13}$CO and C$^{18}$O data have been fitted simultaneously.
 	}
    \label{fig:cornerplots_west}
\end{figure*}

\begin{figure*}[ht!]
\centering
	\includegraphics[scale=0.3]{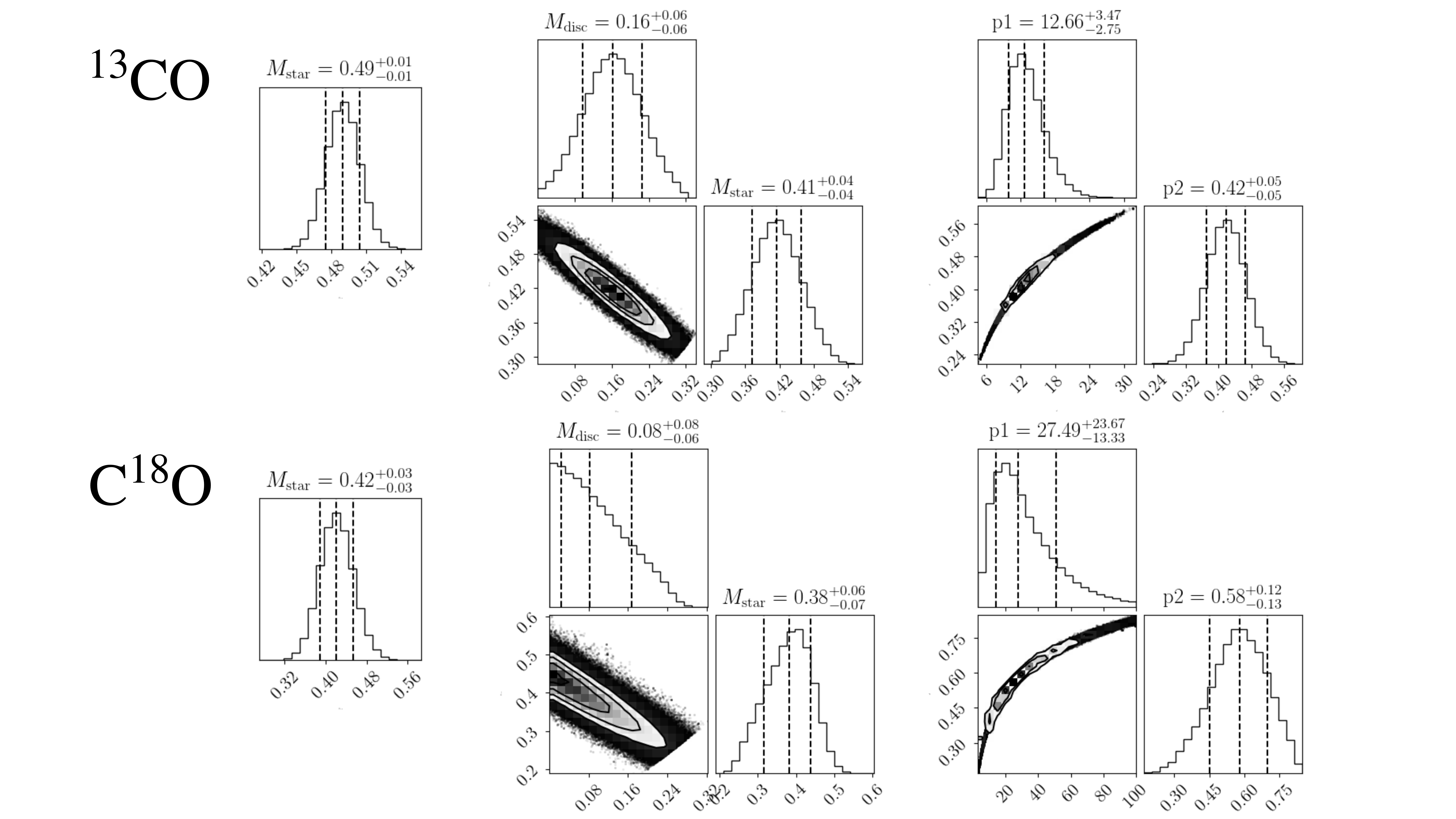}
 	\caption{Probability distribution and corner plots obtained with the Keplerian (left), disk+star (centre) and power law (right) model for the West side, where the $^{13}$CO (upper panels) and C$^{18}$O (lower panels) data have been fitted separately.
 	}
    \label{fig:cornerplots_west_sep}
\end{figure*}

Fig.~\ref{fig:cornerplots_west} collects the probability distribution and corner plots obtained with the Keplerian (left), self-gravitating (disk+star, centre) and power-law (right) fit of the CO isotopologues velocity data, when fitting the West side, simultaneously for the $^{13}$CO and C$^{18}$O isotopologues.  Fig.~\ref{fig:cornerplots_west_sep} shows the same results but for a fit where the two CO-isotopologues (top: $^{13}$CO; bottom: C$^{18}$O) have been fitted separately. 

The fit results correspond to those discussed in Sec.~\ref{sec:singleside}. The final masses and errors are computed from the median value, 16th and 84th percentile uncertainties derived from the posteriors. 

\bibliography{biblio}{}

\begin{thebibliography}{}
\expandafter\ifx\csname natexlab\endcsname\relax\def\natexlab#1{#1}\fi
\providecommand{\url}[1]{\href{#1}{#1}}

\bibitem[{{Andrews} {et~al.}(2018){Andrews}, {Terrell}, {Tripathi}, {Ansdell},
  {Williams}, \& {Wilner}}]{andrews18a}
{Andrews}, S.~M., {Terrell}, M., {Tripathi}, A., {et~al.} 2018, ArXiv e-prints,
  arXiv:1808.10510

\bibitem[{{Andrews} {et~al.}(2009){Andrews}, {Wilner}, {Hughes}, {Qi}, \&
  {Dullemond}}]{andrews09}
{Andrews}, S.~M., {Wilner}, D.~J., {Hughes}, A.~M., {Qi}, C., \& {Dullemond},
  C.~P. 2009, \apj, 700, 1502

\bibitem[{{Ansdell} {et~al.}(2016){Ansdell}, {Williams}, {van der Marel},
  {Carpenter}, {Guidi}, {Hogerheijde}, {Mathews}, {Manara}, {Miotello},
  {Natta}, {Oliveira}, {Tazzari}, {Testi}, {van Dishoeck}, \& {van
  Terwisga}}]{ansdell16a}
{Ansdell}, M., {Williams}, J.~P., {van der Marel}, N., {et~al.} 2016, \apj,
  828, 46

\bibitem[{{Bae} \& {Zhu}(2018)}]{bae18b}
{Bae}, J., \& {Zhu}, Z. 2018, \apj, 859, 119

\bibitem[{{Barbieri} {et~al.}(2005){Barbieri}, {Fraternali}, {Oosterloo},
  {Bertin}, {Boomsma}, \& {Sancisi}}]{barbieri05}
{Barbieri}, C.~V., {Fraternali}, F., {Oosterloo}, T., {et~al.} 2005, \aap, 439,
  947

\bibitem[{{Bergin} \& {Williams}(2018)}]{bergin18a}
{Bergin}, E.~A., \& {Williams}, J.~P. 2018, arXiv e-prints, arXiv:1807.09631

\bibitem[{{Bergin} {et~al.}(2013){Bergin}, {Cleeves}, {Gorti}, {Zhang},
  {Blake}, {Green}, {Andrews}, {Evans}, {Henning}, {{\"O}berg}, {Pontoppidan},
  {Qi}, {Salyk}, \& {van Dishoeck}}]{bergin13}
{Bergin}, E.~A., {Cleeves}, L.~I., {Gorti}, U., {et~al.} 2013, \nat, 493, 644

\bibitem[{Bertin \& Lodato(1999)}]{bertin99}
Bertin, G., \& Lodato, G. 1999, A\&A, 350, 694

\bibitem[{{Bosman} {et~al.}(2018){Bosman}, {Walsh}, \& {van
  Dishoeck}}]{bosman18}
{Bosman}, A.~D., {Walsh}, C., \& {van Dishoeck}, E.~F. 2018, \aap, 618, A182

\bibitem[{{Cadman} {et~al.}(2020){Cadman}, {Hall}, {Rice}, {Harries}, \&
  {Klaassen}}]{cadman20}
{Cadman}, J., {Hall}, C., {Rice}, K., {Harries}, T.~J., \& {Klaassen}, P.~D.
  2020, \mnras, 498, 4256

\bibitem[{{Calcino} {et~al.}(2020){Calcino}, {Christiaens}, {Price}, {Pinte},
  {Davis}, {van der Marel}, \& {Cuello}}]{calcino20}
{Calcino}, J., {Christiaens}, V., {Price}, D.~J., {et~al.} 2020, \mnras, 498,
  639

\bibitem[{{Cleeves} {et~al.}(2018){Cleeves}, {{\"O}berg}, {Wilner}, {Huang},
  {Loomis}, {Andrews}, \& {Guzman}}]{cleeves18}
{Cleeves}, L.~I., {{\"O}berg}, K.~I., {Wilner}, D.~J., {et~al.} 2018, \apj,
  865, 155

\bibitem[{{Cossins} {et~al.}(2009){Cossins}, {Lodato}, \& {Clarke}}]{CLC09}
{Cossins}, P., {Lodato}, G., \& {Clarke}, C.~J. 2009, MNRAS, 393, 1157

\bibitem[{{Dipierro} {et~al.}(2014){Dipierro}, {Lodato}, {Testi}, \& {de
  Gregorio Monsalvo}}]{dipierro14}
{Dipierro}, G., {Lodato}, G., {Testi}, L., \& {de Gregorio Monsalvo}, I. 2014,
  MNRAS, 444, 1919

\bibitem[{{Dong} {et~al.}(2015){Dong}, {Hall}, {Rice}, \& {Chiang}}]{dong15}
{Dong}, R., {Hall}, C., {Rice}, K., \& {Chiang}, E. 2015, ApJ, 812, L32

\bibitem[{{Draine}(2003)}]{draine03}
{Draine}, B.~T. 2003, \araa, 41, 241

\bibitem[{{Favre} {et~al.}(2013){Favre}, {Cleeves}, {Bergin}, {Qi}, \&
  {Blake}}]{favre13}
{Favre}, C., {Cleeves}, L.~I., {Bergin}, E.~A., {Qi}, C., \& {Blake}, G.~A.
  2013, \apjl, 776, L38

\bibitem[{{Flaherty} {et~al.}(2020){Flaherty}, {Hughes}, {Simon}, {Qi}, {Bai},
  {Bulatek}, {Andrews}, {Wilner}, \& {K{\'o}sp{\'a}l}}]{flaherty20}
{Flaherty}, K., {Hughes}, A.~M., {Simon}, J.~B., {et~al.} 2020, \apj, 895, 109

\bibitem[{Foreman-Mackey(2016)}]{corner}
Foreman-Mackey, D. 2016, The Journal of Open Source Software, 1, 24.
\newblock \url{https://doi.org/10.21105/joss.00024}

\bibitem[{{Foreman-Mackey} {et~al.}(2013){Foreman-Mackey}, {Hogg}, {Lang}, \&
  {Goodman}}]{emcee}
{Foreman-Mackey}, D., {Hogg}, D.~W., {Lang}, D., \& {Goodman}, J. 2013, \pasp,
  125, 306

\bibitem[{{Forgan} {et~al.}(2018){Forgan}, {Ilee}, \& {Meru}}]{forgan18a}
{Forgan}, D.~H., {Ilee}, J.~D., \& {Meru}, F. 2018, \apjl, 860, L5

\bibitem[{{Gaia Collaboration} {et~al.}(2018){Gaia Collaboration}, {Brown},
  {Vallenari}, {Prusti}, {de Bruijne}, {Babusiaux}, {Bailer-Jones}, {Biermann},
  {Evans}, {Eyer}, {Jansen}, {Jordi}, {Klioner}, {Lammers}, {Lindegren},
  {Luri}, {Mignard}, {Panem}, {Pourbaix}, {Randich}, {Sartoretti}, {Siddiqui},
  {Soubiran}, {van Leeuwen}, {Walton}, {Arenou}, {Bastian}, {Cropper},
  {Drimmel}, {Katz}, {Lattanzi}, {Bakker}, {Cacciari}, {Casta{\~n}eda},
  {Chaoul}, {Cheek}, {De Angeli}, {Fabricius}, {Guerra}, {Holl}, {Masana},
  {Messineo}, {Mowlavi}, {Nienartowicz}, {Panuzzo}, {Portell}, {Riello},
  {Seabroke}, {Tanga}, {Th{\'e}venin}, {Gracia-Abril}, {Comoretto},
  {Garcia-Reinaldos}, {Teyssier}, {Altmann}, {Andrae}, {Audard},
  {Bellas-Velidis}, {Benson}, {Berthier}, {Blomme}, {Burgess}, {Busso},
  {Carry}, {Cellino}, {Clementini}, {Clotet}, {Creevey}, {Davidson}, {De
  Ridder}, {Delchambre}, {Dell'Oro}, {Ducourant},
  {Fern{\'a}ndez-Hern{\'a}ndez}, {Fouesneau}, {Fr{\'e}mat}, {Galluccio},
  {Garc{\'\i}a-Torres}, {Gonz{\'a}lez-N{\'u}{\~n}ez}, {Gonz{\'a}lez-Vidal},
  {Gosset}, {Guy}, {Halbwachs}, {Hambly}, {Harrison}, {Hern{\'a}ndez},
  {Hestroffer}, {Hodgkin}, {Hutton}, {Jasniewicz}, {Jean-Antoine-Piccolo},
  {Jordan}, {Korn}, {Krone-Martins}, {Lanzafame}, {Lebzelter}, {L{\"o}ffler},
  {Manteiga}, {Marrese}, {Mart{\'\i}n-Fleitas}, {Moitinho}, {Mora}, {Muinonen},
  {Osinde}, {Pancino}, {Pauwels}, {Petit}, {Recio-Blanco}, {Richards},
  {Rimoldini}, {Robin}, {Sarro}, {Siopis}, {Smith}, {Sozzetti}, {S{\"u}veges},
  {Torra}, {van Reeven}, {Abbas}, {Abreu Aramburu}, {Accart}, {Aerts},
  {Altavilla}, {{\'A}lvarez}, {Alvarez}, {Alves}, {Anderson}, {Andrei},
  {Anglada Varela}, {Antiche}, {Antoja}, {Arcay}, {Astraatmadja}, {Bach},
  {Baker}, {Balaguer-N{\'u}{\~n}ez}, {Balm}, {Barache}, {Barata}, {Barbato},
  {Barblan}, {Barklem}, {Barrado}, {Barros}, {Barstow}, {Bartholom{\'e}
  Mu{\~n}oz}, {Bassilana}, {Becciani}, {Bellazzini}, {Berihuete}, {Bertone},
  {Bianchi}, {Bienaym{\'e}}, {Blanco-Cuaresma}, {Boch}, {Boeche}, {Bombrun},
  {Borrachero}, {Bossini}, {Bouquillon}, {Bourda}, {Bragaglia}, {Bramante},
  {Breddels}, {Bressan}, {Brouillet}, {Br{\"u}semeister}, {Brugaletta},
  {Bucciarelli}, {Burlacu}, {Busonero}, {Butkevich}, {Buzzi}, {Caffau},
  {Cancelliere}, {Cannizzaro}, {Cantat-Gaudin}, {Carballo}, {Carlucci},
  {Carrasco}, {Casamiquela}, {Castellani}, {Castro-Ginard}, {Charlot},
  {Chemin}, {Chiavassa}, {Cocozza}, {Costigan}, {Cowell}, {Crifo}, {Crosta},
  {Crowley}, {Cuypers}, {Dafonte}, {Damerdji}, {Dapergolas}, {David}, {David},
  {de Laverny}, {De Luise}, {De March}, {de Martino}, {de Souza}, {de Torres},
  {Debosscher}, {del Pozo}, {Delbo}, {Delgado}, {Delgado}, {Di Matteo},
  {Diakite}, {Diener}, {Distefano}, {Dolding}, {Drazinos}, {Dur{\'a}n},
  {Edvardsson}, {Enke}, {Eriksson}, {Esquej}, {Eynard Bontemps}, {Fabre},
  {Fabrizio}, {Faigler}, {Falc{\~a}o}, {Farr{\`a}s Casas}, {Federici},
  {Fedorets}, {Fernique}, {Figueras}, {Filippi}, {Findeisen}, {Fonti},
  {Fraile}, {Fraser}, {Fr{\'e}zouls}, {Gai}, {Galleti}, {Garabato},
  {Garc{\'\i}a-Sedano}, {Garofalo}, {Garralda}, {Gavel}, {Gavras}, {Gerssen},
  {Geyer}, {Giacobbe}, {Gilmore}, {Girona}, {Giuffrida}, {Glass}, {Gomes},
  {Granvik}, {Gueguen}, {Guerrier}, {Guiraud}, {Guti{\'e}rrez-S{\'a}nchez},
  {Haigron}, {Hatzidimitriou}, {Hauser}, {Haywood}, {Heiter}, {Helmi}, {Heu},
  {Hilger}, {Hobbs}, {Hofmann}, {Holland}, {Huckle}, {Hypki}, {Icardi},
  {Jan{\ss}en}, {Jevardat de Fombelle}, {Jonker}, {Juh{\'a}sz}, {Julbe},
  {Karampelas}, {Kewley}, {Klar}, {Kochoska}, {Kohley}, {Kolenberg},
  {Kontizas}, {Kontizas}, {Koposov}, {Kordopatis}, {Kostrzewa-Rutkowska},
  {Koubsky}, {Lambert}, {Lanza}, {Lasne}, {Lavigne}, {Le Fustec}, {Le
  Poncin-Lafitte}, {Lebreton}, {Leccia}, {Leclerc}, {Lecoeur-Taibi},
  {Lenhardt}, {Leroux}, {Liao}, {Licata}, {Lindstr{\o}m}, {Lister}, {Livanou},
  {Lobel}, {L{\'o}pez}, {Managau}, {Mann}, {Mantelet}, {Marchal}, {Marchant},
  {Marconi}, {Marinoni}, {Marschalk{\'o}}, {Marshall}, {Martino}, {Marton},
  {Mary}, {Massari}, {Matijevi{\v{c}}}, {Mazeh}, {McMillan}, {Messina},
  {Michalik}, {Millar}, {Molina}, {Molinaro}, {Moln{\'a}r}, {Montegriffo},
  {Mor}, {Morbidelli}, {Morel}, {Morris}, {Mulone}, {Muraveva}, {Musella},
  {Nelemans}, {Nicastro}, {Noval}, {O'Mullane}, {Ord{\'e}novic},
  {Ord{\'o}{\~n}ez-Blanco}, {Osborne}, {Pagani}, {Pagano}, {Pailler},
  {Palacin}, {Palaversa}, {Panahi}, {Pawlak}, {Piersimoni}, {Pineau}, {Plachy},
  {Plum}, {Poggio}, {Poujoulet}, {Pr{\v{s}}a}, {Pulone}, {Racero}, {Ragaini},
  {Rambaux}, {Ramos-Lerate}, {Regibo}, {Reyl{\'e}}, {Riclet}, {Ripepi}, {Riva},
  {Rivard}, {Rixon}, {Roegiers}, {Roelens}, {Romero-G{\'o}mez}, {Rowell},
  {Royer}, {Ruiz-Dern}, {Sadowski}, {Sagrist{\`a} Sell{\'e}s}, {Sahlmann},
  {Salgado}, {Salguero}, {Sanna}, {Santana-Ros}, {Sarasso}, {Savietto},
  {Schultheis}, {Sciacca}, {Segol}, {Segovia}, {S{\'e}gransan}, {Shih},
  {Siltala}, {Silva}, {Smart}, {Smith}, {Solano}, {Solitro}, {Sordo}, {Soria
  Nieto}, {Souchay}, {Spagna}, {Spoto}, {Stampa}, {Steele},
  {Steidelm{\"u}ller}, {Stephenson}, {Stoev}, {Suess}, {Surdej}, {Szabados},
  {Szegedi-Elek}, {Tapiador}, {Taris}, {Tauran}, {Taylor}, {Teixeira},
  {Terrett}, {Teyssand ier}, {Thuillot}, {Titarenko}, {Torra Clotet}, {Turon},
  {Ulla}, {Utrilla}, {Uzzi}, {Vaillant}, {Valentini}, {Valette}, {van Elteren},
  {Van Hemelryck}, {van Leeuwen}, {Vaschetto}, {Vecchiato}, {Veljanoski},
  {Viala}, {Vicente}, {Vogt}, {von Essen}, {Voss}, {Votruba}, {Voutsinas},
  {Walmsley}, {Weiler}, {Wertz}, {Wevers}, {Wyrzykowski}, {Yoldas},
  {{\v{Z}}erjal}, {Ziaeepour}, {Zorec}, {Zschocke}, {Zucker}, {Zurbach}, \&
  {Zwitter}}]{gaia18}
{Gaia Collaboration}, {Brown}, A.~G.~A., {Vallenari}, A., {et~al.} 2018, \aap,
  616, A1

\bibitem[{{Gradshteyn} \& {Ryzhik}(1980)}]{gradshteyn80}
{Gradshteyn}, I.~S., \& {Ryzhik}, I.~M. 1980, {Table of integrals, series and
  products}

\bibitem[{{Hall} {et~al.}(2018){Hall}, {Rice}, {Dipierro}, {Forgan}, {Harries},
  \& {Alexander}}]{halletal2018}
{Hall}, C., {Rice}, K., {Dipierro}, G., {et~al.} 2018, \mnras, 477, 1004

\bibitem[{{Hall} {et~al.}(2020){Hall}, {Dong}, {Teague}, {Terry}, {Pinte},
  {Paneque-Carre{\~n}o}, {Veronesi}, {Alexander}, \& {Lodato}}]{hall20}
{Hall}, C., {Dong}, R., {Teague}, R., {et~al.} 2020, \apj, 904, 148

\bibitem[{{Huang} {et~al.}(2018{\natexlab{a}}){Huang}, {Andrews}, {P{\'e}rez},
  {Zhu}, {Dullemond}, {Isella}, {Benisty}, {Bai}, {Birnstiel}, {Carpenter},
  {Guzm{\'a}n}, {Hughes}, {{\"O}berg}, {Ricci}, {Wilner}, \&
  {Zhang}}]{huang18c}
{Huang}, J., {Andrews}, S.~M., {P{\'e}rez}, L.~M., {et~al.} 2018{\natexlab{a}},
  \apjl, 869, L43

\bibitem[{{Huang} {et~al.}(2018{\natexlab{b}}){Huang}, {Andrews}, {Dullemond},
  {Isella}, {P{\'e}rez}, {Guzm{\'a}n}, {{\"O}berg}, {Zhu}, {Zhang}, {Bai},
  {Benisty}, {Birnstiel}, {Carpenter}, {Hughes}, {Ricci}, {Weaver}, \&
  {Wilner}}]{huang18b}
{Huang}, J., {Andrews}, S.~M., {Dullemond}, C.~P., {et~al.} 2018{\natexlab{b}},
  \apjl, 869, L42

\bibitem[{{Huang} {et~al.}(2020){Huang}, {Andrews}, {{\"O}berg}, {Ansdell},
  {Benisty}, {Carpenter}, {Isella}, {P{\'e}rez}, {Ricci}, {Williams}, {Wilner},
  \& {Zhu}}]{huang20}
{Huang}, J., {Andrews}, S.~M., {{\"O}berg}, K.~I., {et~al.} 2020, \apj, 898,
  140

\bibitem[{{Hunter}(2007)}]{matplotlib}
{Hunter}, J.~D. 2007, Computing in Science and Engineering, 9, 90

\bibitem[{{Ilee} {et~al.}(2017){Ilee}, {Forgan}, {Evans}, {Hall}, {Booth},
  {Clarke}, {Rice}, {Boley}, {Caselli}, {Hartquist}, \& {Rawlings}}]{ilee17}
{Ilee}, J.~D., {Forgan}, D.~H., {Evans}, M.~G., {et~al.} 2017, \mnras, 472, 189

\bibitem[{{Kratter} \& {Lodato}(2016)}]{kratter16}
{Kratter}, K., \& {Lodato}, G. 2016, \araa, 54, 271

\bibitem[{{Lodato} \& {Bertin}(2003)}]{lodato03}
{Lodato}, G., \& {Bertin}, G. 2003, \aap, 398, 517

\bibitem[{{Lodato} \& {Rice}(2004)}]{lodato04}
{Lodato}, G., \& {Rice}, W.~K.~M. 2004, \mnras, 351, 630

\bibitem[{{Long} {et~al.}(2017){Long}, {Herczeg}, {Pascucci}, {Drabek-Maunder},
  {Mohanty}, {Testi}, {Apai}, {Hendler}, {Henning}, {Manara}, \&
  {Mulders}}]{long17a}
{Long}, F., {Herczeg}, G.~J., {Pascucci}, I., {et~al.} 2017, \apj, 844, 99

\bibitem[{{Luhman} \& {Rieke}(1999)}]{luhman99}
{Luhman}, K.~L., \& {Rieke}, G.~H. 1999, \apj, 525, 440

\bibitem[{{Macias} {et~al.}(2021){Macias}, {Guerra-Alvarado},
  {Carrasco-Gonzalez}, {Ribas}, {Espaillat}, {Huang}, \& {Andrews}}]{macias21}
{Macias}, E., {Guerra-Alvarado}, O., {Carrasco-Gonzalez}, C., {et~al.} 2021,
  arXiv e-prints, arXiv:2102.04648

\bibitem[{{Meru} {et~al.}(2017){Meru}, {Juh{\'a}sz}, {Ilee}, {Clarke},
  {Rosotti}, \& {Booth}}]{meruetal2017}
{Meru}, F., {Juh{\'a}sz}, A., {Ilee}, J.~D., {et~al.} 2017, \apjl, 839, L24

\bibitem[{{Miotello} {et~al.}(2016){Miotello}, {van Dishoeck}, {Kama}, \&
  {Bruderer}}]{miotello16a}
{Miotello}, A., {van Dishoeck}, E.~F., {Kama}, M., \& {Bruderer}, S. 2016,
  \aap, 594, A85

\bibitem[{{Miotello} {et~al.}(2017){Miotello}, {van Dishoeck}, {Williams},
  {Ansdell}, {Guidi}, {Hogerheijde}, {Manara}, {Tazzari}, {Testi}, {van der
  Marel}, \& {van Terwisga}}]{miotello17}
{Miotello}, A., {van Dishoeck}, E.~F., {Williams}, J.~P., {et~al.} 2017, \aap,
  599, A113

\bibitem[{{Paneque-Carreno} {et~al.}(2021){Paneque-Carreno}, {Perez},
  {Benisty}, {Hall}, {Veronesi}, {Lodato}, {Sierra}, {Carpenter}, {Andrews},
  {Bae}, {Henning}, {Kwon}, {Linz}, {Loinard}, {Pinte}, {Ricci}, {Tazzari},
  {Testi}, \& {Wilner}}]{paneque20}
{Paneque-Carreno}, T., {Perez}, L.~M., {Benisty}, M., {et~al.} 2021, arXiv
  e-prints, arXiv:2103.14048

\bibitem[{{Pascucci} {et~al.}(2016){Pascucci}, {Testi}, {Herczeg}, {Long},
  {Manara}, {Hendler}, {Mulders}, {Krijt}, {Ciesla}, {Henning}, {Mohanty},
  {Drabek-Maunder}, {Apai}, {Sz{\H u}cs}, {Sacco}, \& {Olofsson}}]{pascucci16a}
{Pascucci}, I., {Testi}, L., {Herczeg}, G.~J., {et~al.} 2016, \apj, 831, 125

\bibitem[{{P{\'e}rez} {et~al.}(2016){P{\'e}rez}, {Carpenter}, {Andrews},
  {Ricci}, {Isella}, {Linz}, {Sargent}, {Wilner}, {Henning}, {Deller},
  {Chandler}, {Dullemond}, {Lazio}, {Menten}, {Corder}, {Storm}, {Testi},
  {Tazzari}, {Kwon}, {Calvet}, {Greaves}, {Harris}, \& {Mundy}}]{perezelias227}
{P{\'e}rez}, L.~M., {Carpenter}, J.~M., {Andrews}, S.~M., {et~al.} 2016,
  Science, 353, 1519

\bibitem[{{Pinte} {et~al.}(2018){Pinte}, {M{\'e}nard}, {Duch{\^e}ne}, {Hill},
  {Dent}, {Woitke}, {Maret}, {van der Plas}, {Hales}, {Kamp}, {Thi}, {de
  Gregorio-Monsalvo}, {Rab}, {Quanz}, {Avenhaus}, {Carmona}, \&
  {Casassus}}]{pinte18}
{Pinte}, C., {M{\'e}nard}, F., {Duch{\^e}ne}, G., {et~al.} 2018, \aap, 609, A47

\bibitem[{{Pinte} {et~al.}(2020){Pinte}, {Price}, {M{\'e}nard}, {Duch{\^e}ne},
  {Christiaens}, {Andrews}, {Huang}, {Hill}, {van der Plas}, {Perez}, {Isella},
  {Boehler}, {Dent}, {Mentiplay}, \& {Loomis}}]{pinte20}
{Pinte}, C., {Price}, D.~J., {M{\'e}nard}, F., {et~al.} 2020, \apjl, 890, L9

\bibitem[{Pringle(1981)}]{pringle81}
Pringle, J.~E. 1981, ARA\&A, 19, 137

\bibitem[{{Qu{\'e}nard} {et~al.}(2018){Qu{\'e}nard}, {Jim{\'e}nez-Serra},
  {Viti}, {Holdship}, \& {Coutens}}]{quenard18}
{Qu{\'e}nard}, D., {Jim{\'e}nez-Serra}, I., {Viti}, S., {Holdship}, J., \&
  {Coutens}, A. 2018, \mnras, 474, 2796

\bibitem[{Rice {et~al.}(2004)Rice, Lodato, Pringle, Armitage, \&
  Bonnell}]{rice04}
Rice, W. K.~M., Lodato, G., Pringle, J.~E., Armitage, P.~J., \& Bonnell, I.~A.
  2004, MNRAS, 355, 543

\bibitem[{Rice {et~al.}(2006)Rice, Lodato, Pringle, Armitage, \&
  Bonnell}]{rice06}
---. 2006, MNRAS, 372, L9

\bibitem[{{Trapman} {et~al.}(2017){Trapman}, {Miotello}, {Kama}, {van
  Dishoeck}, \& {Bruderer}}]{trapman17}
{Trapman}, L., {Miotello}, A., {Kama}, M., {van Dishoeck}, E.~F., \&
  {Bruderer}, S. 2017, \aap, 605, A69

\bibitem[{{van der Marel} {et~al.}(2020){van der Marel}, {Birnstiel}, {Garufi},
  {Ragusa}, {Christiaens}, {Price}, {Sallum}, {Muley}, {Francis}, \&
  {Dong}}]{marel20}
{van der Marel}, N., {Birnstiel}, T., {Garufi}, A., {et~al.} 2020, arXiv
  e-prints, arXiv:2010.10568

\bibitem[{{Veronesi} {et~al.}(2019){Veronesi}, {Lodato}, {Dipierro}, {Ragusa},
  {Hall}, \& {Price}}]{veronesi19}
{Veronesi}, B., {Lodato}, G., {Dipierro}, G., {et~al.} 2019, \mnras, 489, 3758

\bibitem[{{Williams} \& {Best}(2014)}]{williams14a}
{Williams}, J.~P., \& {Best}, W.~M.~J. 2014, \apj, 788, 59

\bibitem[{{Zhang} {et~al.}(2018){Zhang}, {Zhu}, {Huang}, {Guzm{\'a}n},
  {Andrews}, {Birnstiel}, {Dullemond}, {Carpenter}, {Isella}, {P{\'e}rez},
  {Benisty}, {Wilner}, {Baruteau}, {Bai}, \& {Ricci}}]{zhang18}
{Zhang}, S., {Zhu}, Z., {Huang}, J., {et~al.} 2018, \apjl, 869, L47

\end{thebibliography}
\bibliographystyle{aasjournal}

%\end{thebibliography}

\end{document}